\documentclass[final]{IEEEtran}
\usepackage[dvips]{graphics}
\usepackage{epsf,rotating,color,graphicx,float,amssymb,amsmath}
\usepackage{epsfig,url,times,cite,subfigure,soul}
\usepackage{booktabs}
\usepackage{multirow}
\usepackage{lipsum}
\usepackage{algorithm,algorithmic}

\usepackage{calc}
\usepackage{epstopdf}
\newenvironment{notationlist}[1]
{\begin{list}{}%
		{\renewcommand\makelabel[1]{##1\hfill}%
			\settowidth\labelwidth{\makelabel{#1}}%
			\setlength\leftmargin{\labelwidth+\labelsep}
			\setlength\itemsep{0\baselineskip}}}%
	{\end{list}}

\newcommand{\define}{\newcommand}
\define{\be}{\begin{eqnarray}}
\define{\ee}{\end{eqnarray}}
\define{\ben}{\begin{eqnarray*}}
	\define{\een}{\end{eqnarray*}}

\begin{document}

\title{\large Chance-Constrained Day-Ahead Hourly Scheduling in Distribution System Operation}

\author{\large Yi Gu$^1$, \textit{Student Member}, IEEE, Huaiguang Jiang$^2$, \textit{Member}, IEEE, Jun Jason Zhang$^1$, \textit{Senior Member}, IEEE, Yingchen Zhang$^2$, \textit{Senior Member}, IEEE, Eduard Muljadi$^2$, \textit{Fellow}, IEEE, and Francisco J. Solis$^3$, \textit{Senior Member}

  {\normalsize $^1$Dept.~of Electrical and Computer Engineering, University of Denver, Denver, CO, 80210\\
               $^2$National Renewable Energy Laboratory, Golden, CO, 80401\\
               $^2$School of Mathematical and Natural Sciences, Arizona State University, Glendale, AZ, 85306\\}
}




\date{}
\maketitle

\begin{abstract}
	
	This paper aims to propose a two-step approach for day-ahead hourly scheduling in a distribution system operation, which contains two operation costs, the operation cost at substation level and feeder level. In the first step, the objective is to minimize the electric power purchase from the day-ahead market with the stochastic optimization. The historical data of day-ahead hourly electric power consumption is used to provide the forecast results with the forecasting error, which is presented by a chance constraint and formulated into a deterministic form by Gaussian mixture model (GMM). In the second step, the objective is to minimize the system loss. Considering the nonconvexity of the three-phase balanced AC optimal power flow problem in distribution systems, the second-order cone program (SOCP) is used to relax the problem. Then, a distributed optimization approach is built based on the alternating direction method of multiplier (ADMM). The results shows that the validity and effectiveness method.

\end{abstract}


	\section*{Nomenclature}
\begin{notationlist}{MAXIMUMSPC}
	
	\item[$C$]  Total cost of the two steps.
	
	\item[$f_{1}$] Operation cost at the substation level.
	
	\item[$f_{2}$]  System loss at feeder levels.
	
	\item[$\beta$] The system loss weight.
	
	\item[$c^{DA},c^{RT}, c^{PV}$] The unit price of the day-ahead market, real-time market and renewable generation.
	
	\item[$c^{s}$] The price to resell the redundant power generated to the electric market.
	
	\item[$G^{DA}$] The electric power obtained from the day-ahead market.
	
	\item[$G^{RT}$] The RT power consuming. 
	
	\item[$G^{R}$] The renewable generation.
	
	
	\item[$\lambda_t$] An occurrence probability helps to decide if the systems needs to buy the electric power at time $t$.
	
	\item[$G_t^{DL}$] The demand load at period t.
	
	
	\item[$P_{ij}$] The loss of branch lines from bus $i$ to $j$.
	
	\item[$z_{ij}$] The impedance from bus $i$ to $j$.
	
	
	
	\item[$G_{err}$]  Forecasting error model of the renewable energy.
	
	
	\item[$V_i$] The voltage at bus $i$.
	
	
	
	
	\item[$n$] The amount of the clusters in the model, $n=1,\cdots,N$.
	
	\item[$\epsilon$] The weight of each cluster in GMM.
	
	\item[$x$] $x = [x_1, x_2, \cdots, x_q]^T$ is the $q$-dimensional data vector.
	
	
	
	
	
	
	
	
	\item [$U_i, C_i$] The parent node and the children nodes at node $i$, $i\in\mathcal{N}_B$.
	
	
	
	\item [$p_i,q_i$]  Active and reactive power at bus $i$.
	
	\item [$\mathbf{V}_i,\mathbf{I}_i$]  The voltage and the current.
	
	\item[$s_i$]  Injection power at bus $i$.
	
	\item[$\Omega$]  Complex impedance.
	
	\item[$G_f$] The forecasted renewable generation.
	
	
	
\end{notationlist}
\section{Background and Motivation}\label{sec:introduction}

In general power system operation, the day-ahead hourly scheduling~\cite{amjady2006day}supplies the customers playing a more active role. They can be provided more benefits, such as lower system operation cost, advanced system reliability, and lower volatility in hourly~\cite{wu2013hourly}. The high penetration of the renewable energy~\cite{gu2014statistical} can lead to a lower net load power consumption from the bulk power system, however, it brings increasingly stochastic deviations in net load profiles~\cite{jiang2016short}. Compared with the transmission systems, the distribution systems often work in an unbalanced polyphase state because of the asynchrony, asymmetry and the diversity of the load, which brings bigger challenge for the distribution system operation. In this paper, a stochastic optimization based two-step approach for day-ahead scheduling is proposed to minimize the distribution system operation cost, which consists of the cost of net load consumption and the system loss.

At the substation level, the high penetration of the renewable energy can reduce the electric power~\cite{carrasco2006power} purchasing from the day-ahead market for the distribution systems~\cite{wu2014chance}. In~\cite{baker2017energy, wallace2005applications, ruszczynski1993parallel}, the stochastic programming optimization (SPO) can provide many potential benefits to the transmission systems. With the renewable energies, this paper focuses on minimizing the electric power purchasing from the day-ahead market at the substation level in the first step. The historical data of the day-ahead electric power purchasing is used to generate the forecast results and the forecasting errors, which can be formulated as a chance constraint for the SPO in distribution systems. According to the distribution of the forecasting errors, the chance constraint can be formulated into a deterministic form by Gaussian Mixture Model (GMM)~\cite{huang2005gaussian}, and the global minimum of the convex problem can be determined. 

In~\cite{wu2013hourly},the variability of the renewable energy~\cite{jiang2014synchrophasor} is managed by hourly demand response in day-ahead scheduling, without considering the stochastic net load deviation in an hour, which dramatically impacts the operation cost of system loss. However, the single line model is used compute the system loss, which ignores the three-phase balanced configuration in distribution systems. In the second step, this paper focuses on minimizing the cost of the system loss at the feeder level with the three-phase balanced model. The characteristic of the nonconvexity is a critical problem for the three-phase balanced AC optimal power flow. The heuristic methods are always applied to solve the problem, but which are hardly avoided falling into the local minimums~\cite{tang2010globallypso}. Based on the alternating direction methods of multipliers (ADMM), a distributed method is provided to solve the AC optimal power flow problem.

This paper is organized as follows: the proposed approach is described in Section~\ref{sec:model}. In Section~\ref{sec:sub}, the operation cost in substation level is analyzed. In Section~\ref{sec:feeder}, the operation cost in feeder level is analyzed. In Section~\ref{sec:num}, the numerical results are presented in IEEE standard distribution systems. The conclusion is Section~\ref{sec:con}.

\section{Architecture and Summary of the Proposed Approach}\label{sec:model}

\begin{figure*}[!t]
	\begin{center}
		\includegraphics [width=1.3\columnwidth, angle=90]{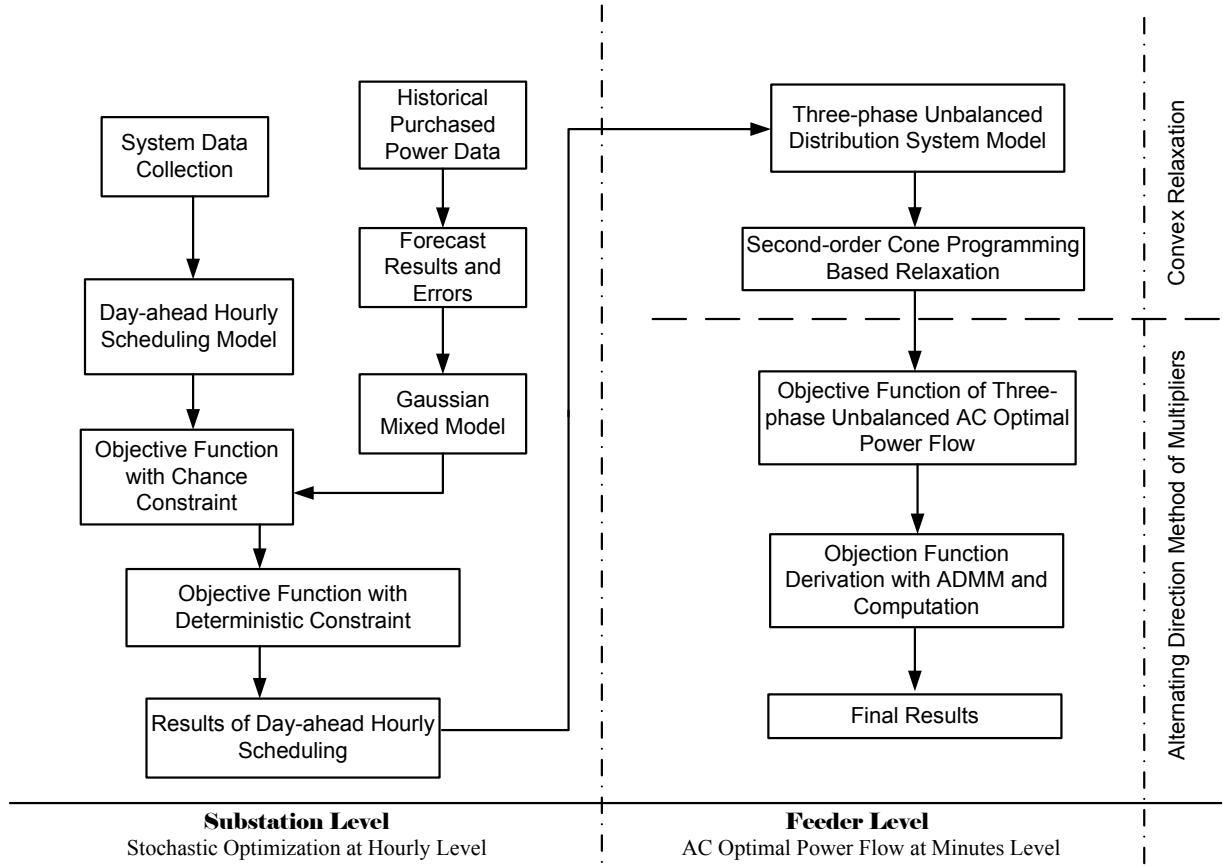}
		\caption{The flowchart of proposed approach.}
		\label{fig:flowchart}
	\end{center}
\end{figure*}

In Fig.~\ref{fig:flowchart}, the proposed method consists two steps, the stochastic optimization of the electric power purchased from the day-ahead market~\cite{conejo2005day} at the substation level and the system loss optimization for the three-phase balanced AC optimal power flow~\cite{cheng1995three,moghaddas2009distributed} at the feeder level~\cite{gu2016knowledge}. 

 Fig.~\ref{fig:flowchart} shows the first step at the left, the historical data of the electric power purchased from the day-ahead market is employed to generate the forecast results and the distribution of the forecast errors. Then, combined with the day-ahead hourly scheduling model, an objective formulation is  simulated with an chance constraint. A GMM based approach is used to convert the chance constraint into a deterministic problem. Finally, the day-ahead hourly optimal operation cost can be obtained at the first step.

In the rest side of Fig.~\ref{fig:flowchart}, a three-phase balanced AC optimal power flow is simulated to compute the distribution system at the feeder level. After the day-ahead hourly purchased electric power is determined, the three-phase balanced distribution system model is built to minimum the system loss. Considering the nonconvexity of the AC optimal power flow, an inequality constraint is built based on SOCP to relax the problem into a convex problem. After that, the objective function of system loss can be derived with ADMM. Finally, the three-phase balanced AC optimal power flow is used to minimize the system loss successfully.

\section{Operation Cost at the Substation Level}~\label{sec:sub}

The total cost is calculated as (\ref{eq:eq01}), the sum of the operation cost at the substation level and feeder level:
\be\label{eq:eq01}
C=\min(f_{1}+\beta f_{2})\
\ee
where $C$ is the total cost, $f_1$ is the purchased electric power from the day-ahead market, which equals to the operation cost at the substation level, $f_2$ is the system loss of the distribution system, which equals to the operation cost at feeder level. $\beta$ is used to limit the system loss weight in the optimization.The consymption of the load cost model at the substation level is simulated as:

\begin{equation}\label{eq:eq02}
\begin{aligned}
f_{1}=&\sum_{t=1}^{NT}\Big\{(c^{DA}G_t^{DA}+{c}^{PV}G_{t}^{PV})+(\lambda_t\cdot{c_t^{RT}}G_t^{RT})\\
&+\big[(1-\lambda_t)\cdot{c^s}(G_t^{DA}+G_t^{PV}-G^{DL}_t)\big]\Big\}
\end{aligned}
\end{equation}

where the time intervals is defined as $t=1,2,…,NT$. $c^{DA}$, $c^{RT}$ and $c^{PV}$ describe the unit price in the day-ahead, real-time market and solar generation, respectively. $c^s$ is the established price to sell the redundant power generated by distribution systems to the electric market, $G^{DA}$ and $G^{RT}$ are the electric power amount purchased from the day-ahead and real time market, $G^{PV}$ is the amount of the solar power generation in the distribution systems. $\lambda_t$ is an occurrence probability which decides if the distribution systems needs to buy the electric from  real-time market at time $t$. $G^{DL}_t$ is the demand load at time $t$. $c^{DA}G_t^{DA}+{c}^{PV}G_{t}^{PV}$ is the base-case generation cost consists of the time-dependent power consumption by day-ahead scheduling and the renewable energy generation cost. $\lambda_t\cdot{c_t^{RT}}G_t^{RT}$ presents the deviation power purchased from real-time market for compensation. And $(1-\lambda_t)\cdot{c^s}(G_t^{DA}+G_t^{PV}-G^{DL}_t)$ means the redundant energy can be resold to the market in a lower price.

Several months of hourly power forecasting data in day-ahead scheduling and actual net load power consumption are used to provide the forecast results~\cite{jiang2016short, yang2017short, jiang2017short }. The available hourly forecasting power in day-ahead market $G^{DA}(NT+1)$ can be now described as following ($G_{err}$ is the day-ahead forecasting error)~\cite{atwa2010optimal}:
\be\label{eq:eq03}
G^{DA}(NT+1)=G^{DA}(NT)+G_{err}(NT)\cdot{G^{DA}(NT)}
\ee

Subject to:
\begin{subequations}
	\be\label{equ:eps1}
	G^{DA}+\lambda{G^{RT}}+G^{PV}=G^{DL}
	\ee
	
	\be\label{equ:eps2}
	c^s<{c^{DA}<c_t^{RT}}
	\ee
	
	\be\label{equ:eps3}
	G^{RT,min}\leq G^{RT}\leq G^{RT,max}\\
	G^{DA,min}\leq G^{DA}\leq G^{DA,max}\\
	G^{PV,min}\leq G^{PV}\leq G^{PV,max}
	\ee
	
	\be\label{equ:eps4}
	\lambda_t =\begin{cases}1 & G^{DA}_t+G^{PV}_{t} < G^{DL}_t\\0 & G^{DA}_t+G^{PV}_{t}  \geq G^{DL}_t\end{cases} 
	\ee
	
	
	\be\label{equ:eps6}
	Pr(f(G_{err})\leq0)> \alpha
	\ee
\end{subequations}
where (\ref{equ:eps6}) indicates that day-ahead forecasting error~\cite{taylor2006comparison} should be fulfilled with the probability $\alpha$, and the chance-constraint~\cite{jana2004stochastic} can be converted in a deterministic formulation with the GMM.

\subsection{GMM with Expectation Maximization. }

Comparing with the regular GMM, a definition of expectation maximization EM based GMM  is described. It is used to model the forecasting error model of the renewable generation. We have known that GMM is a particular form of the finite mixture model. For (\ref{GMM_1}), more than one components  is calculated as the sum with different weights $(\epsilon)$:

\begin{equation}\label{GMM_1}
p(x|\mathbf{\Theta)}=\epsilon{p}(x|\mathbf{\theta})
\end{equation}

$\epsilon$ in (\ref{GMM_1}) is calculated in~\cite{peng2017model} and has been described to be non-negative, that the sum equals to $1$. Each component in the GMM model is a normal distribution and obeys to $\mathbf{\theta}_n=(\mathbf{\mu}, \mathbf{\Sigma})$, which indicates the means vector and the covariance. 

According to~\cite{moon1996expectation}, the parameters of the mixture model is computed with expectation maximization (EM), the expectation-step (called E-step) and maximization-step (called M-step) are described as below.

\subsubsection{E-Step}
 The algorithm is ended when the function in (\ref{GMM_2}) reaches the convergence.

\begin{equation}\label{GMM_2}
R(\theta|\theta_t)=E_{x,\theta_t}[logL(\theta;x)]
\end{equation}

Where $L(\theta;x)= p(x|\theta)$. $x$ is a set of the observed data from the given statistic model and the $\theta$ is unknown parameters along with the likelihood function in~(\ref{GMM_2}). 



\subsubsection{M-Step}

In M-step, the parameters are recalculated and estimated to maximize the quantity of the expectation in~(\ref{GMM_8}).

\begin{equation}\label{GMM_8}
\theta_{(t+1)}=\arg\max R(\theta|\theta_t)
\end{equation}




The EM based GMM can decide the amount of clusters based on minimum description length (MDL)~\cite{tenmoto1998mdl, liang1992parameter}. The MDL criterion is frequently used on the field of selection in~\cite{mclachlan2004finite}, because of the improved method is less sensitive to the initialization.

\section{Operation Cost at the Feeder Level}\label{sec:feeder}
When the substation level cost is determined, the objective function of the three-phase balanced AC optimal power flow is to minimum the system loss. As we know, the influence of the three-phase distribution system is small enough, the second order cone programming (SOCP) is used to calculate the system loss here~\cite{jabr2006radial}.  which can be defined as follows~\cite{pe78ng2014distributed,low2014co78nvex}:
\begin{equation}\label{eq:eq04}
F = \sum_{\mathcal{E}}P_{ij}, 
\end{equation}
where $P_{ij}$ is a branch loss from bus $i$ to $j$, and $P_{ij}$ = $|I_{ij}|^2 r_{ij}$, $I_{ij}$ is the complex current from bus $i$ to $j$, and $z_{ij}$ is the complex impedance $z_{ij} = r_{ij} + \textbf{i} x_{ij}$. $\mathcal{E}$ is the branch set of the distribution system, which can be represented with the set of buses and branches: $\mathcal{G} = [\mathcal{V}, \mathcal{E}]$.  

Based on the branch flow model, the SOCP relaxation inequalities can be represented as follows
\begin{equation}
\frac{|S_{ij}|^2}{v_i} \leq l_{ij}, \label{equ:eps44}
\end{equation}
where $S_{ij}$ indicate the complex power flow $S_{ij} = P_{ij} + \textbf{i} Q_{ij}$, $l_{ij}:=|I_{ij}|^2$, and $v_i:=|V_i|^2$.
The basic physical constraints can be defined as follows:
\begin{subequations}
	\begin{align} 
    V_{i,min} \leq V_i \leq V_{i,max}, \label{equ:eps66} \\
    I_{ij} \leq I_{ij,max}. \label{equ:eps7} 
	\end{align}
\end{subequations}

The standard optimization problem of ADMM is defined as follows:
\begin{subequations}
	\begin{align} 
    &\min_{x,z} \ \ f(x)+g(z)   \label{equ:eps8}\\
    &\ s.t.  \ \  x \in \mathcal{K}_{x}, z\in\mathcal{K}_z \label{equ:eps9}\\
    &\        \ \  \ \ \     Ax +Bz=c  \nonumber
	\end{align}
\end{subequations}
 
Then, a dual problem is build based on the objective function (\ref{eq:eq04}) and (\ref{equ:eps8}) with ADMM and solved in parallel.


\section{The Numerical Results}\label{sec:num}
\begin{figure}[h!]
	\begin{center}
		\subfigure[]{ \label{fig:IEEE 123}
		     \includegraphics[width=1.0\columnwidth,height=4.0cm, angle=0]{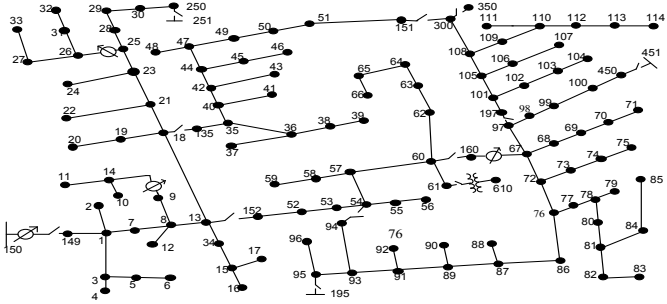}}
		\subfigure[]{ \label{fig:Error Curve}
		      \includegraphics[width=1.0\columnwidth,height=4.0cm, angle=0]{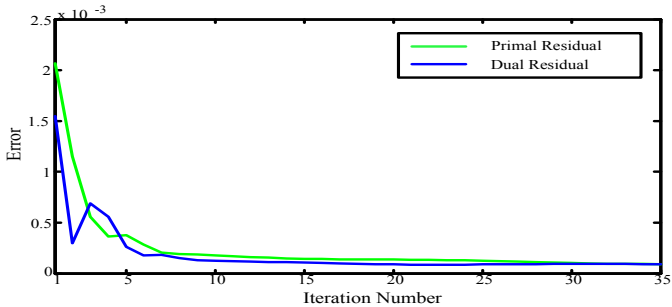}}
		\caption{(a) The IEEE 123-bus distribution system. (b) The residual error curves of the three-phase balanced AC optimal power flow.}\label{fig:signal-analysis}
\end{center}
\end{figure}

\begin{table}[]
	\centering
	\caption{Time consuming of different algorithms based on three individual systems}\vspace{-0.00in}
	\label{comparison_1}
	\begin{tabular}{|l|l|l|l|}
		\hline
		Method & IEEE 13-bus& IEEE 34-bus& IEEE 123-bus\\ \hline
		Proposed Method & 1.78 s & 5.23 s & 14.66 s \\ \hline
		Simulated annealing& 189.43 s & 221.34 s & 459.21 s \\ \hline
		Interior-Point & 7.20 s & 13.16 s & 25.58 s \\ \hline
		
	\end{tabular}
\end{table}

\subsection{The test bench}
As shown in Fig.~\ref{fig:IEEE 123}, the IEEE 123-bus distribution system contains 118 basic branches, 85 unbalanced loads, 4 capacitors, and 11 three phase switches (6 initially closed and 5 initially opened), which is the topology taken in the preliminary results. The simulation platform is based on a computer server with a Xeon processor and 32 GB ram. The programming language are Python and Matlab.

\subsection{The performance of the proposed approach}
It is assumed that the renewable energy resources are located in bus 7, 23, 29, 35, 47, 49, 65, 76, 83, and 99~\cite{quezada2006assessment}. The result of the three-phase balanced AC optimal power flow is shown in Fit.~\ref{fig:Error Curve}. The errors of the primal residual and the dual residual are less than $0.5 \times 10^{-3}$ after 5 iterations. After 30 iterations (less than 0.2 second), the curves of primal residual and the dual residual are coinciding and stable, which demonstrates the high speed and effectiveness of the convergency of the proposed approach. 

\begin{figure}[!t]
\begin{center}
		\includegraphics [width=1.1\columnwidth,height=4.5cm, angle=0]{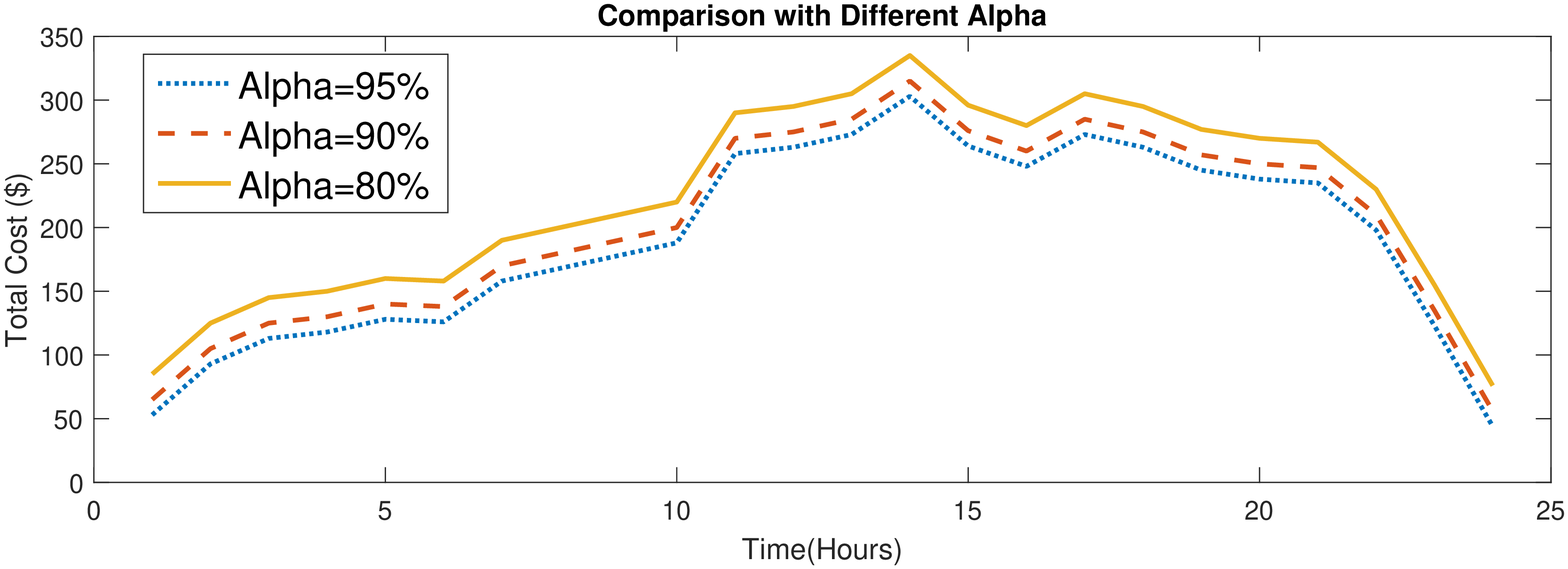}
		\caption{The total cost with different Alpha.}\label{fig:Alpha}
\end{center}
\end{figure}

In Fig.~\ref{fig:Alpha}, the comparison is made with different $\alpha$ in~(\ref{equ:eps6}) during 24 hours. The results shows that the total operation cost of the system is higher with a lower $\alpha$, which indicates that a higher accuracy of error forecasting model can help to save more money.

\subsection{Performance comparison}
As in Table~\ref{comparison_1}, the different algorithms are used on different individual power systems (IEEE trans 13, -34 and -123 Bus ). The proposed approach obtains a shortest time consuming on all of the three power systems, which demonstrates our method can work more efficiently than others.

\section{Conclusion}~\label{sec:con}
In this paper, a stochastic optimization based approach is proposed for the chance-constrained day-ahead hourly scheduling problem in distribution system operation. The operation cost is divided into two parts, the operation cost at substation level and feeder level. In the operation cost at substation level, the proposed approach minimizes the electric power purchase from the day-ahead market with a stochastic optimization. In the operation cost at feeder level, the system loss is presented with the three-phase balanced AC optimal power flow~\cite{banavar2015overview}. In our work, the detailed flowchart and the description of the stochastic optimization with the forecasting errors is improved, which helps to describe our approach detailedly. And the detailed description of the derivation from the chance constraint into the deterministic form with GMM is provided. The SOCP relaxation is used to solve the problem with three-phase balanced distribution system. 

In the future, we will focus on improving an advanced approach to develop the distribution system optimization based on an three-phased unbalanced system~\cite{lin1999three, kersting2001radial, khushalani2007development, jiang2017big}.


\bibliographystyle{IEEEbib}
\bibliography{refs_liu_2017_asilomar}

\begin{thebibliography}{10}

\bibitem{amjady2006day}
Nima Amjady,
\newblock ``Day-ahead price forecasting of electricity markets by a new fuzzy
  neural network,''
\newblock {\em IEEE Transactions on power systems}, vol. 21, no. 2, pp.
  887--896, 2006.

\bibitem{wu2013hourly}
Hongyu Wu, Mohammad Shahidehpour, and Ahmed Al-Abdulwahab,
\newblock ``Hourly demand response in day-ahead scheduling for managing the
  variability of renewable energy,''
\newblock {\em IET Generation, Transmission \& Distribution}, vol. 7, no. 3,
  pp. 226--234, 2013.

\bibitem{gu2014statistical}
Yi~Gu, Huaiguang Jiang, Yingchen Zhang, and David~Wenzhong Gao,
\newblock ``Statistical scheduling of economic dispatch and energy reserves of
  hybrid power systems with high renewable energy penetration,''
\newblock in {\em Signals, Systems and Computers, 2014 48th Asilomar Conference
  on}. IEEE, 2014, pp. 530--534.

\bibitem{jiang2016short}
Huaiguang Jiang, Yingchen Zhang, Eduard Muljadi, Jun Zhang, and Wenzhong Gao,
\newblock ``A short-term and high-resolution distribution system load
  forecasting approach using support vector regression with hybrid parameters
  optimization,''
\newblock {\em IEEE Transactions on Smart Grid}, 2016.

\bibitem{carrasco2006power}
Juan~Manuel Carrasco, Leopoldo~Garcia Franquelo, Jan~T Bialasiewicz, Eduardo
  Galv{\'a}n, Ram{\'o}n~Carlos PortilloGuisado, MA~Martin Prats,
  Jos{\'e}~Ignacio Le{\'o}n, and Narciso Moreno-Alfonso,
\newblock ``Power-electronic systems for the grid integration of renewable
  energy sources: A survey,''
\newblock {\em IEEE Transactions on industrial electronics}, vol. 53, no. 4,
  pp. 1002--1016, 2006.

\bibitem{wu2014chance}
Hongyu Wu, Mohammad Shahidehpour, Zuyi Li, and Wei Tian,
\newblock ``Chance-constrained day-ahead scheduling in stochastic power system
  operation,''
\newblock {\em IEEE Transactions on Power Systems}, vol. 29, no. 4, pp.
  1583--1591, 2014.

\bibitem{baker2017energy}
Kyri Baker, Gabriela Hug, and Xin Li,
\newblock ``Energy storage sizing taking into account forecast uncertainties
  and receding horizon operation,''
\newblock {\em IEEE Transactions on Sustainable Energy}, vol. 8, no. 1, pp.
  331--340, 2017.

\bibitem{wallace2005applications}
Stein~W Wallace and William~T Ziemba,
\newblock {\em Applications of stochastic programming},
\newblock SIAM, 2005.

\bibitem{ruszczynski1993parallel}
Andrzej Ruszczy{\'n}ski,
\newblock ``Parallel decomposition of multistage stochastic programming
  problems,''
\newblock {\em Mathematical programming}, vol. 58, no. 1, pp. 201--228, 1993.

\bibitem{huang2005gaussian}
Yonghong Huang, Kevin~B Englehart, Bernard Hudgins, and Adrian~DC Chan,
\newblock ``A gaussian mixture model based classification scheme for
  myoelectric control of powered upper limb prostheses,''
\newblock {\em IEEE Transactions on Biomedical Engineering}, vol. 52, no. 11,
  pp. 1801--1811, 2005.

\bibitem{jiang2014synchrophasor}
Huaiguang Jiang, Jun~Jason Zhang, David~Wenzhong Gao, Yingchen Zhang, and
  Eduard Muljadi,
\newblock ``Synchrophasor based auxiliary controller to enhance power system
  transient voltage stability in a high penetration renewable energy
  scenario,''
\newblock in {\em Power Electronics and Machines for Wind and Water
  Applications (PEMWA), 2014 IEEE Symposium}. IEEE, 2014, pp. 1--7.

\bibitem{tang2010globallypso}
Zaiyong Tang and Kallol~Kumar Bagchi,
\newblock ``Globally convergent particle swarm optimization via
  branch-and-bound,''
\newblock {\em Computer and Information Science}, vol. 3, no. 4, pp. 60--71,
  2010.

\bibitem{conejo2005day}
Antonio~J Conejo, Miguel~A Plazas, Rosa Espinola, and Ana~B Molina,
\newblock ``Day-ahead electricity price forecasting using the wavelet transform
  and arima models,''
\newblock {\em IEEE transactions on power systems}, vol. 20, no. 2, pp.
  1035--1042, 2005.

\bibitem{cheng1995three}
Carol~S Cheng and Dariush Shirmohammadi,
\newblock ``A three-phase power flow method for real-time distribution system
  analysis,''
\newblock {\em IEEE Transactions on Power Systems}, vol. 10, no. 2, pp.
  671--679, 1995.

\bibitem{moghaddas2009distributed}
SM~Moghaddas-Tafreshi and Elahe Mashhour,
\newblock ``Distributed generation modeling for power flow studies and a
  three-phase unbalanced power flow solution for radial distribution systems
  considering distributed generation,''
\newblock {\em Electric Power Systems Research}, vol. 79, no. 4, pp. 680--686,
  2009.

\bibitem{gu2016knowledge}
Yi~Gu, Huaiguang Jiang, Yingchen Zhang, Jun~Jason Zhang, Tianlu Gao, and Eduard
  Muljadi,
\newblock ``Knowledge discovery for smart grid operation, control, and
  situation awarenessÑa big data visualization platform,''
\newblock in {\em North American Power Symposium (NAPS), 2016}. IEEE, 2016, pp.
  1--6.

\bibitem{yang2017short}
Rui Yang, Huaiguang Jiang, and Yingchen Zhang,
\newblock ``Short-term state forecasting-based optimal voltage regulation in
  distribution systems: Preprint,''
\newblock Tech. {R}ep., NREL (National Renewable Energy Laboratory (NREL),
  Golden, CO (United States)), 2017.

\bibitem{jiang2017short}
Huaiguang Jiang, Fei Ding, Yingchen Zhang, Huaiguang Jiang, Fei Ding, and
  Yingchen Zhang,
\newblock ``Short-term load forecasting based automatic distribution network
  reconfiguration: Preprint,''
\newblock Tech. {R}ep., National Renewable Energy Laboratory (NREL), Golden, CO
  (United States), 2017.

\bibitem{atwa2010optimal}
YM~Atwa, EF~El-Saadany, MMA Salama, and R~Seethapathy,
\newblock ``Optimal renewable resources mix for distribution system energy loss
  minimization,''
\newblock {\em IEEE Transactions on Power Systems}, vol. 25, no. 1, pp.
  360--370, 2010.

\bibitem{taylor2006comparison}
James~W Taylor, Lilian~M De~Menezes, and Patrick~E McSharry,
\newblock ``A comparison of univariate methods for forecasting electricity
  demand up to a day ahead,''
\newblock {\em International Journal of Forecasting}, vol. 22, no. 1, pp.
  1--16, 2006.

\bibitem{jana2004stochastic}
RK~Jana* and MP~Biswal,
\newblock ``Stochastic simulation-based genetic algorithm for chance constraint
  programming problems with continuous random variables,''
\newblock {\em International Journal of Computer Mathematics}, vol. 81, no. 9,
  pp. 1069--1076, 2004.

\bibitem{peng2017model}
Weishi Peng,
\newblock ``Model selection for gaussian mixture model based on desirability
  level criterion,''
\newblock {\em Optik-International Journal for Light and Electron Optics}, vol.
  130, pp. 797--805, 2017.

\bibitem{moon1996expectation}
Todd~K Moon,
\newblock ``The expectation-maximization algorithm,''
\newblock {\em IEEE Signal processing magazine}, vol. 13, no. 6, pp. 47--60,
  1996.

\bibitem{tenmoto1998mdl}
Hiroshi Tenmoto, Mineichi Kudo, and Masaru Shimbo,
\newblock ``Mdl-based selection of the number of components in mixture models
  for pattern classification,''
\newblock {\em Advances in Pattern Recognition}, pp. 831--836, 1998.

\bibitem{liang1992parameter}
Zhengrong Liang, Ronald~J Jaszczak, and R~Edward Coleman,
\newblock ``Parameter estimation of finite mixtures using the em algorithm and
  information criteria with application to medical image processing,''
\newblock {\em IEEE Transactions on Nuclear Science}, vol. 39, no. 4, pp.
  1126--1133, 1992.

\bibitem{mclachlan2004finite}
Geoffrey McLachlan and David Peel,
\newblock {\em Finite mixture models},
\newblock John Wiley \& Sons, 2004.

\bibitem{jabr2006radial}
Rabih~A Jabr,
\newblock ``Radial distribution load flow using conic programming,''
\newblock {\em IEEE transactions on power systems}, vol. 21, no. 3, pp.
  1458--1459, 2006.

\bibitem{pe78ng2014distributed}
Qiuyu Peng and Steven~H Low,
\newblock ``Distributed algorithm for optimal power flow on a radial network,''
\newblock in {\em 2014 IEEE 53rd Annual Conference on Decision and Control
  (CDC)}. IEEE, 2014, pp. 167--172.

\bibitem{low2014co78nvex}
Steven~H Low,
\newblock ``Convex relaxation of optimal power flowÑpart i: Formulations and
  equivalence,''
\newblock {\em IEEE Transactions on Control of Network Systems}, vol. 1, no. 1,
  pp. 15--27, 2014.

\bibitem{quezada2006assessment}
VH~M{\'e}ndez Quezada, J~Rivier Abbad, and T~Gomez~San Roman,
\newblock ``Assessment of energy distribution losses for increasing penetration
  of distributed generation,''
\newblock {\em IEEE Transactions on power systems}, vol. 21, no. 2, pp.
  533--540, 2006.

\bibitem{banavar2015overview}
Mahesh~K Banavar, Jun~J Zhang, Bhavana Chakraborty, Homin Kwon, Ying Li,
  Huaiguang Jiang, Andreas Spanias, Cihan Tepedelenlioglu, Chaitali
  Chakrabarti, and Antonia Papandreou-Suppappola,
\newblock ``An overview of recent advances on distributed and agile sensing
  algorithms and implementation,''
\newblock {\em Digital Signal Processing}, vol. 39, pp. 1--14, 2015.

\bibitem{lin1999three}
Whei-Min Lin, Yuh-Sheng Su, Hong-Chan Chin, and Jen-Hao Teng,
\newblock ``Three-phase unbalanced distribution power flow solutions with
  minimum data preparation,''
\newblock {\em IEEE Transactions on power Systems}, vol. 14, no. 3, pp.
  1178--1183, 1999.

\bibitem{kersting2001radial}
William~H Kersting,
\newblock ``Radial distribution test feeders,''
\newblock in {\em Power Engineering Society Winter Meeting, 2001. IEEE}. IEEE,
  2001, vol.~2, pp. 908--912.

\bibitem{khushalani2007development}
Sarika Khushalani, Jignesh~M Solanki, and Noel~N Schulz,
\newblock ``Development of three-phase unbalanced power flow using pv and pq
  models for distributed generation and study of the impact of dg models,''
\newblock {\em IEEE Transactions on Power Systems}, vol. 22, no. 3, pp.
  1019--1025, 2007.

\bibitem{jiang2017big}
Huaiguang Jiang, Yan Li, Yingchen Zhang, Jun~Jason Zhang, David~Wenzhong Gao,
  Eduard Muljadi, and Yi~Gu,
\newblock ``Big data-based approach to detect, locate, and enhance the
  stability of an unplanned microgrid islanding,''
\newblock {\em Journal of Energy Engineering}, vol. 143, no. 5, pp. 04017045,
  2017.

\end{thebibliography}

\end{document}